\documentstyle[12pt]{article}
\makeatletter
\@addtoreset{equation}{section}
\makeatother

\begin{document}
\newcommand{\COM}[1]{[[[#1]]]}
\newcommand{\ot}{\frac{1}{2}}
\newcommand{\D}{& \displaystyle}  
\newcommand{\di}{\displaystyle}   
\font \math=msym10 scaled \magstep 1
\font \mathi=msym10 scaled \magstep 0
\def\eps{{\epsilon}}
\def\myr{{\mbox{\math R}}}
\def\myri{{\mbox{\mathi R}}}
\def\myc{{\mbox{\math C}}}
\def\myx{{\mbox{\math X}}}
\def\myn{{\mbox{\math N}}}
\date{October 1992}
\title
{The Mechanism of Complex Langevin Simulations\thanks{Supported by Fonds
zur F\"orderung der Wissenschaftlichen
Forschung in \"Osterreich, project P7849.} }
\author
{{\bf H. Gausterer}\\ \\
Institut f\"ur Theoretische Physik,\\
Universit\"at Graz, A-8010 Graz, AUSTRIA\\ \\
{\bf Sean Lee}\\ \\
Department of Physics,\\
University of Florida,
Gainesville, Florida, USA}
\maketitle
\begin{abstract}
We discuss conditions under which expectation values computed from a complex
Langevin process $Z$ will converge to integral averages over a given complex
valued weight function. The difficulties in proving a general result
are pointed out. For complex valued polynomial actions, it is shown that
for a process converging to a strongly stationary process
one gets the correct answer for averages of polynomials if
$ c_{\tau}(k) \equiv E(e^{ikZ(\tau)}) $ satisfies certain conditions.
If these conditions are not satisfied, then the stochastic process
is not necessarily described by a complex Fokker Planck equation.
The result is illustrated
with the exactly solvable complex frequency harmonic oscillator.
\end{abstract}
\newpage

\section{Introduction}

It is well known that a straightforward application of standard
simulation techniques like the Metropolis {\it et al}. algorithm \cite{Met}
will fail when they are applied to problems with a complex action or
Hamiltonian. This is due to the fact that there is no direct
probabilistic interpretation of a distribution function of the form
$ e^{-S} $ when the action $ S $ is complex valued. As an alternative to such
algorithms, the complex Langevin (CL) method was first proposed by
Klauder \cite{Klaud1},
and subsequently studied by many authors. The central idea of CL is based on
the fact that for a Langevin equation there is no formal restriction to a real
valued drift term. The use of a complex drift term provides CL with a genuine
 advantage over other methods; namely, since  CL uses the entire complex action
$ S $
to define a stochastic process, it can in principle converge directly to
the desired distribution. This potential for circumventing the well known
``sign problems''  of other standard algorithms is just one reason CL
continues to be a subject of great interest.

 Quite independent of its utility as a numerical
 technique is the interesting fact that under certain conditions a system
governed by a
 complex Hamiltonian can  nevertheless be given a probabilistic interpretation.
Indeed,
 the name complex Langevin  may be slightly misleading, since the Langevin
equations really
 describe a real diffusion  process in twice as many dimensions.

Unfortunately, there is currently no complete theory of the CL method.
Many conditions under which a real Langevin process can be shown to converge to
a given
distribution  are not satisfied for a general CL process. Furthermore, from the
point of view of
 numerical simulations, in some cases CL  is known to converge to the wrong
results \cite{Klaud2,LinHir}.
 For simple actions, this truant behavior can be corrected by an appropriate
choice
 of kernel in the Langevin equation \cite{Schul}, but for more general systems,
in particular
 lattice fermion models of current interest,  it is far from clear which choice
of
 kernel is required.

The purpose of this paper is to explore in a rigorous fashion the conditions
under which the CL process correctly simulates a given system with a complex
Hamiltonian
defined on a Euclidean space $ x \in \myr^n $. While a general theorem is still
lacking, we will
demonstrate a set of sufficient conditions for polynomial
actions.

\section{The Process}
Throughout the paper we will assume that the system is described by
variables $x \in \myr^n$ with some complex action or Hamiltonian
$ S : \myr^n \rightarrow \myc$,  where $\mbox{Re} S$ is bounded
from below. For simplicity the discussion will be restricted to one
degree of freedom, since the following statements allow for an immediate
generalization to $\myr^n$. The quantities of physical interest
are of the form
\begin{equation}
\langle g(x) \rangle = {1 \over \cal{N} }\int_\myri g(x) e^{-S(x)} dx,
\end{equation}
\begin{equation}
{\cal{N}} = \int_\myri e^{-S(x)} dx \;\;\;,
\end{equation}
assuming that for the partition function $\cal{N}$ we have
 $0<|\cal{N}|< \infty$.
Both $g(z)$ and $S(z)$ are assumed to be analytic in $\myc$.
Since $S(z)$ is analytic, this provides a local Lipschitz condition
for the Langevin equation \ref{1}; that is, \ref{1} has a unique
local solution
that  is defined up to a random explosion
time \cite{Arn}.
For real actions  $S \in \myr$ we define a
process $\{X(\tau), \tau \geq 0\}$ by the Langevin equation
\begin{equation}
dX(\tau) = F(X(\tau)) d\tau + dW(\tau) \;\;,
\end{equation}
with the drift term
\begin{equation}
F(x) =-{1 \over 2} {{dS(x)} \over {dx}} \;\;,
\end{equation}
where $W(\tau)$ is a standard Wiener process with zero mean and cova\-riance
\begin{equation}
<W(\tau_1)W(\tau_2)> = \mbox{min} (\tau_1,\tau_2).
\end{equation}
The probability density $ f(x, \tau) $ for such a process will
 converge pointwise to the
desired distribution function. That is
\begin{equation}
\lim_{\tau \rightarrow \infty} f(x,\tau) = \hat f(x)
\mbox{ a.e.}\;\;,
\end{equation}
with
\begin{equation}
\hat f(x) = {1 \over \cal{N} } e^{-S(x)} \;\; .
\end{equation}

As already mentioned in the introduction,
 for the complex case one can formally construct a Langevin equation
\begin{equation}\label{1}
dZ(\tau) = F(Z(\tau)) d\tau + dW(\tau) \;\;,
\end{equation}
with the drift term
\begin{equation}
F(z) =-{1 \over 2} {{dS(z)} \over {dz}} \;\;.
\end{equation}
As above, $W(\tau)$ is a standard Wiener process.
This is actually a two dimensional process of the form
\begin{equation} \label{7}
dX(\tau) = G(X(\tau),Y(\tau)) d\tau + dW(\tau)
\end{equation}
\begin{equation}\label{8}
dY(\tau) = H(X(\tau),Y(\tau)) d\tau
\end{equation}
with $S(z)=u(x,y) + iv(x,y)$
\begin{equation}
G(x,y) =-{1 \over 2} {{\partial u(x,y)} \over {\partial x}} \;, \;\;\;
H(x,y) = {1 \over 2} {{\partial u(x,y)} \over {\partial y}} \;\;.
\end{equation}
Note that the equation for $dY$ has a zero diffusion coefficient,
 but is nevertheless a stochastic equation through its dependence on $X$ .

The process $\{(X(\tau),Y(\tau)), \tau \geq 0\}$
as defined by equation \ref{1}
has a distribution density $f(x,y,\tau)$. There are
now two crucial questions concerning this process. The first question
is whether this so defined process converges in distribution
at all to some
$(X,Y)$,
\begin{equation}
\lim_{\tau \rightarrow \infty} f(x,y,\tau) = \hat f(x,y),
\mbox{ a.e.} \;\;.
\end{equation}
The second question has to do with the problem of whether
$\hat f(x,y)$ satisfies
\begin{equation}\label{2}
E(g(X+ i Y)) = \int_{\myri^2} g(x+iy) \hat f(x,y) dxdy =
{1 \over \cal{N} }\int_\myri g(x) e^{-S(x)} dx.
\end{equation}
This equation contains the essence of complex Langevin. It tells us
that if the process as defined above  has converged in
distribution, we might be able to calculate $\langle g(x) \rangle$ by an
equivalent
probabilistic system of twice as many dimensions.

Before we investigate equation \ref{2} it is necessary
 to discuss the asymptotic
behavior of $X(\tau),Y(\tau)$  for $\tau \rightarrow \infty$. To do this
we will examine the equivalent Fokker-Planck (F-P) equation. First note
that since the diffusion and drift coefficients are independent of
$\tau$, the process $(X(\tau),Y(\tau))$ is a homogeneous diffusion process.
For twice continuously differentiable distribution densities
with respect to $x,y$ and once with respect to $\tau$,
there exists a F-P equation
\cite{Arn}. Then the equivalent F-P equation for the above Langevin equation
is
\begin{equation}\label{3}
{{\partial f(x,y,\tau)} \over {\partial \tau}} = T f(x,y,\tau)
\end{equation}
with
\begin{equation}
T =
 -G_x(x,y) - G(x,y) { \partial \over { \partial x}}
 -H_y(x,y) - H(x,y) { \partial \over { \partial y}}
 +{ 1\over 2}{{\partial^2}\over {{\partial x}^2}} \;\;.
\end{equation}
As may be seen from the above equation, this case only requires a continuous
 first order derivative with respect to $y$.
Let us first assume that $T$
has a unique stationary solution $\hat f(x,y)$
\begin{equation}
T \hat f(x,y) = 0 \;\;,
\end{equation}
with
\begin{equation}
\hat f(x,y) \geq 0, \mbox{ a.e.}
\end{equation}
and
\begin{equation}
\int_{\myri^2} \hat f(x,y) dxdy = 1 \;\;.
\end{equation}
One can then use $\hat f(x,y)$ to define an invariant
measure
\begin{equation}
\mu(\Omega) = \int_{\Omega} \hat f(x,y) dxdy
\end{equation}
with respect to $L_{\tau}=\exp(Q \tau),\;\;  \tau \geq 0$ , which is
defined as an
operator family
in $L^p(\myr^2,d\mu)$ ($p=1,2$). The operator $Q$ is obtained by transforming
$T$ to $L^1(\myr^2,d\mu)$ by
\begin{equation}
Q = e^{S} T e^{-S} \;\;.
\end{equation}
Then there is a theorem
which tells us that $\{ L_{\tau}, \tau \geq 0 \}$ is a contraction
semigroup \cite{Yos}, from which it follows that
for any function
$\phi \in L^p(\myr^2,d\mu)$ ,
$p=1,2$ one has convergence in the strong sense
\begin{equation}
\mbox{s-}\lim_{\tau \rightarrow \infty}
 L_{\tau} \phi = c_{\phi}\;\;,
\end{equation}
where $c_{\phi}$ is a constant.
Since $ \mu(\myr^2) < \infty$, any distribution
 density $f(x,y,\tau)$ converges pointwise
to the stationary solution
\begin{equation}
\lim_{\tau \rightarrow \infty} f(x,y,\tau) = \hat f(x,y)
\mbox{ a.e.} \;\;.
\end{equation}

On the other hand, if the zero eigenvalue of $T$
is $M$-fold degenerate then there are $M$ ergodic classes.
In this case we have
\begin{equation}
\lim_{\tau \rightarrow \infty} f(x,y,\tau) = \sum_{i=1}^M
c_i \hat f_i(x,y)  \mbox{ a.e.} \;\;.
\end{equation}

Unfortunately, it is still an open question as to what conditions
guarantee the existence of a stationary solution at all.
Note that the
 CL process defined by  the real two dimensional
equations \ref{7} and \ref{8} has a singular
 diffusion matrix.
Therefore, a general
statement on the existence of stationary solutions can not be made based
on the
nature of the drift terms.

Further, $ T $
does not fall into the class of hypoelliptic
operators of constant strength for general $S(x+iy)$,
$(x,y) \in \myr^2$. In this case a general statement on the regularity
of the solutions can not be made \cite{Hoer}, thus one can not exclude
the possibility that the
solutions $T \hat f =0$ exist only
in the sense of distributions (weak solutions). If this is
the case then it might be quite
difficult to find (construct) the stationary density.

It is easy to see that a situation like this can occur
 with the very simple example
$S(x) = c x^2,\;\; c \in \myr^+ \;,\;\; $ which is now supposed to be solved by
complex Langevin. The Langevin equation reads:
\begin{equation}
dX(\tau) = - c X(\tau)d\tau + dW(\tau),
\end{equation}
\begin{equation}
dY(\tau) = -c Y(\tau) d\tau
\end{equation}
The stationary density $\hat f(x,y)$ is then a weak solution and
can be formally given as
\begin{equation}
\hat f(x,y) \sim e^{- c x^2} \delta (y).
\end{equation}

\section{Polynomial Actions}
Returning to the question of equation \ref{2}, we now demonstrate
the following general result:
 let $S(x)$ and $g(x)$ be polynomials of degree $N$ and $M$
($ M \leq N-1$)  and $S(x)$ such that $e^{-S} \in \cal{S}(\myr)$.
$\cal{S}(\myr)$ is the Schwartz space of $C^{\infty}$ functions
of rapid decrease. Assume $Z(\tau)$ converges in distribution
to a strongly stationary process. Then equation \ref{2} holds if
there exists a $\tau_0$ such that for all
$\tau \geq  \tau_0$
\begin{enumerate}
\item
\begin{equation}
\left| E(Z^n(\tau) e^{ikZ(\tau)}) \right| < \infty \mbox{ for all }
0 \leq n \leq N-1, k \in \myr \;\;\;.
\end{equation}
\item The Fourier transform
\begin{equation} \label{10}
h(x,\tau) = {1 \over {2 \pi}} \int_{\myri} c_{\tau}(k) e^{-ikx} dk
\end{equation}
 of the expectation value
\begin{equation} \label{9}
E(e^{ikZ(\tau)}) \equiv c_\tau(k)
\end{equation}
satisfies  $h(x,\tau) \in C^2(\myr)$
with respect to $x$, and $ h(x,t) \in C^1(\myr)$ with respect to $\tau$.
Furthermore, $ x^{N-1} h(x,\tau) \in L^1(\myr,dx) \;\;. $
\item
\begin{equation}\label{11}
 \lim_{ \tau \to \infty } c_{ \tau}(k) \in {\cal S}( \myr)
\end{equation}
\end{enumerate}

Before proving this result, we note that
condition 3 is also a necessary condition for equation {\ref 2},
since, by assumption, $e^{-S} \in {\cal S} $. Also, requiring
$ c_{ \tau}(k) \in {\cal S}( \myr) $ would be a simplification of
the condition 2.  However,
this requirement for all $ c_{\tau}(k) $ might be too restrictive.

Under the above assumptions we have
\begin{equation} \label{6}
\langle e^{ikx(\tau)} \rangle = E(e^{ikZ(\tau)}) \;\; ,
\end{equation}
where $\langle g(x(\tau)) \rangle$  is given by
\begin{equation}
\langle g(x(\tau)) \rangle = \int_{\myri} g(x) h(x,\tau) dx.
\end{equation}

{}From assumption 2 it follows that  $ c_{\tau}(k) \in C^{N-1}( \myr) \;,
$ and  thus  we can conclude that for $f(x) = e^{ikx}$
\begin{equation}
E({ {df(Z(\tau))} \over {dZ(\tau)}} {{dS(Z(\tau))} \over {dZ(\tau)}})
=
\langle { {df(x(\tau))} \over {dx(\tau)}} {{dS(x(\tau))} \over {dx(\tau)}}
\rangle
\end{equation}
\begin{equation}
E({ {d^2f(Z(\tau))} \over {dZ^2(\tau)}})
=
\langle { {d^2f(x(\tau))} \over {dx^2(\tau)}} \rangle \;\;,
\end{equation}
and that the surface terms in the above integral expressions vanish. Thus
$h(x,\tau)$ obeys the pseudo F-P equation with a complex
drift term
\begin{equation}\label{4}
{{\partial h(x,\tau)} \over {\partial \tau}} =
{1 \over 2} {\partial \over {\partial x}} \left[
{{\partial S(x)} \over {\partial x}} + {\partial \over {\partial x}}
\right] h(x,\tau) = \tilde T h(x,\tau).
\end{equation}
The operator $\tilde Q$, which is $\tilde T$ transformed
to $L^1(\myr,d\tilde \mu)$ by
\begin{equation}
\tilde Q = e^{S} \tilde T e^{-S},
\end{equation}
where $d \tilde \mu$ is given by
\begin{equation}
d\tilde \mu(x) = \left| {{1 \over \cal{N}} e^{-S(x)}} \right| dx,
\end{equation}
has a
zero eigenvalue with eigenfunction
\begin{equation}
\hat h(x) = 1 .
\end{equation}
However, we also note that $\tilde Q$
has a second solution of zero eigenvalue in
$L^1(\myr,d\tilde \mu)$ given by
\begin{equation}
\hat h_s(x) = \int^{x}_{x_0} e^{S(y)} dy\;\;.
\end{equation}
Although $e^{-S(x)}\hat h_s(x)$ is in general not positive definite,
and hence automatically
excluded when $ S $ is real, it is not {\it a priori} clear that
$ h(x, \tau) \rightarrow e^{-S(x)}\hat h _s(x) \; \mbox{a.e.}$
is excluded for complex $ S $.

But note
\begin{equation}
e^{-S(x)}\hat h_s(x) = {\cal{O}} ( { 1 \over {x^{N-1} }}) \mbox{ for }
|x|  \rightarrow \infty \;\;,
\end{equation}
where $N$ is the degree of the polynomial $S(x)$.
Thus $ e^{-S(x)}\hat h_s(x) \notin \cal{S}(\myr) \;,$ which contradicts
\ref{11}.
So $ e^{-S(x)}\hat h_s(x) $ cannot be the limit
of $ h(x, \tau)$.
As was true with the real $S$ case, the finite measure
$\tilde \mu$ implies that
all solutions of \ref{4} satisfying \ref{10}
 converge pointwise to the desired
result.

The above 3 conditions appear to be a necessary set which must be satisfied
in order to connect the Langevin process $ Z $ to the complex Fokker Planck
equation \ref{4}. These conditions also guarantee the correct convergence
of the first $ N-1 $ moments. More generally, however, any higher moment
$ E(Z^n(\tau)) $ which does exist will converge correctly if $ h(x,\tau) $
is such that $ \left| \langle x^n(\tau) \rangle \right| <  \infty $.

Two conclusions from this analysis are
especially worth noting. First of all, we have seen that the complex
Fokker-Planck
equation has been derived under certain assumptions about the Langevin process.
If these
criteria are not met, in particular if the function $ c_{\tau}(k) $ is ill
behaved, then
there is no longer necessarily a relation between the CL equation and the
complex Fokker-Planck
equation. This is true even if the spectrum of the complex F-P equation is such
that all solutions
converge to the desired stationary solution.
Secondly, our result implies that the success or failure of the CL method
only depends on the
properties of the two eigenfunctions of the  zero eigenvalue of
$ \tilde Q$
and not on the requirement that
the real part of the spectrum of $\tilde Q$,  $ \mbox{Re} \sigma
(\tilde Q ) \leq 0$ .
Thus, we see that, although it is certainly more convenient
if $ \mbox{Re} \sigma (\tilde Q ) \leq 0$,
an analysis of the spectrum is neither necessary nor
sufficient for studying the behavior
of the CL process.

\section{An Exactly Solvable Example}
 For the Gaussian model $ S(x)= \omega x^2$, $ \omega =a+ib \in \myc^+ $,
 the two dimensional real Fokker-Planck equation has the correct unique
stationary  density to which all initial solutions converge \cite{Hay}.
 The transition density is given by
\begin{equation}
f(x,y, \tau \vert x_0,y_0,0) \sim e^{-r^{T}(\tau) \Sigma ^{-1}( \tau) r(\tau)}
,
\end{equation}
with
\begin{equation}
r_1(\tau)=x-m_1(x_0,y_0,\tau)\;\;,\; r_2(\tau)=y-m_2(x_0,y_0,\tau)\;\;.
\end{equation}
For the detailed form of the matrix $ \Sigma( \tau) $ and
 the vector $m(x_0,y_0,\tau) $, the reader is referred to reference \cite{Hay}.
As mentioned above, the transition density converges to the  stationary
density
\begin{equation}
 \lim_{ \tau \rightarrow \infty} f(x,y,\tau \vert x_0,y_0,0)
\sim \exp [ - ax^2 + 2{a \over b } xy + (1+{{2a^2}\over b^2 } ) y^2 ]
\;\;.
\end{equation}
With the transition density we have
\begin{equation}
c_{\tau}(k) = E(e^{ikZ(\tau)}) = \int_{\myri^2} e^{ikz} dxdy \int_{\myri^2}
f(x,y,\tau \vert x_0, y_0 , 0) f(x_0,y_0,0) dx_0dy_0 \;\;,
\end{equation}
from which it follows that for $ f(x_0,y_0,0) = \delta (x_0 - x^{'}) \delta
 (y_0 - y^{'}) \; $ ,
$ c_{\tau}(k)$ has the form
\begin{equation}
c_{\tau}(k) \sim e^{ -k^2 \alpha(a,b,\tau)} \;\;.
\end{equation}
By examining the form of the matrix $\Sigma(\tau)$, one can see that
there always exists a $\tau_0$ such that for all $ \tau \geq \tau_0$,
the real valued function $ \alpha(a,b,\tau) > 0$.
Clearly, $ c_{\tau}(k) \in \cal S(\myr)$, and  the conditions
1 to 3 hold.

Since in this case the ground state of $T$  is unique, by the
preceeding analysis it follows that all initial states converge to
the proper probability density.

It is interesting to note that for
quadratic polynomials it can be demonstrated that all solutions
to the
complex Fokker-Planck equation which are in
$L^2(\myr,d\tilde \mu)$
converge to the ground state
of $ \tilde Q $ or to zero. Note that all  square integrable solutions
are also absolutely integrable ($L^1$) in this case.

In this example the complex Fokker-Planck equation can be solved
exactly. For the above range of the complex coupling
$\omega$, the spectrum of $\tilde Q$ satisfies  $\mbox{Re}[\sigma(\tilde
 Q)] \leq 0 $ , and the
eigenfunctions of $\tilde Q$ defined on
$L^2(\myr,d\tilde \mu)$
form a basis.

To see this note that
\begin{equation}
\tilde Q = {1\over 2} e^{ S/2} (-H) e^{-S/2} \;\;,
\end{equation}
where $H$ is now the ordinary Schr\"odinger operator for
the complex frequency extension of the harmonic oscillator with
a zero energy groundstate. With the help of the dilatation
operator, which is given by \cite{CyFroKiSi}
\begin{equation}
U_{\theta} \psi(x) = e^{\theta /2} \psi (e^{\theta} x) \;\; ,
\end{equation}
one can now show the completeness of the eigenfunctions of
$H$. First note the domain $D(U_{\theta})$ is dense in $L^2(\myr,dx)$.
Secondly, the eigenfunctions of the real harmonic oscillator are
in $D(U_{\theta})$ for $\mbox{Re}[ \exp(2 \theta)] > 0$. The dilatation
operator maps the eigenfunctions of the real harmonic oscillator to the
eigenfunctions of the complex harmonic oscillator. Now for all
functions $\phi(x) \in D(U_{\theta})$  and $\psi_n(x)$ the
n-th eigenfunction of the real harmonic oscillator we have
\begin{equation}
(\phi, U_{\theta} \psi_n) = (U^{-1}_{\bar{\theta}} \phi, \psi_n) = 0
\;\; \mbox{ , for all } \; n
\end{equation}
if and only if $U^{-1}_{\bar{\theta}} \phi = 0$, which implies
$\phi = 0$. But since $U^{-1}_ {\bar{\theta}} \phi$ for
$\phi \in D(U_{\theta})$ is dense, the eigenfunctions of the
complex harmonic oscillator form a complete basis.
Thus
\begin{equation}
s-\lim_{\tau \rightarrow \infty} h(x,\tau) = \hat h(x) \mbox{ or } 0
\end{equation}
for all solutions $h(x,\tau) \in L^2(\myr,d\tilde \mu)$ of
\begin{equation}
{{ \partial h(x,\tau) } \over { \partial \tau } } = \tilde Q h(x,\tau)
\;\;.
\end{equation}
Again, from $\tilde \mu(\myr) < \infty$ one can conclude
that $ h(x,\tau)$ converges pointwise.

\section{Conclusions}
To date there is no comprehensive theory of the complex
Langevin method. General results are difficult to prove
because many theorems on differential operators and diffusion
processes do not
apply  to $T$ and the Langevin equations. However, for the case
 of polynomial
actions we have demonstrated a set of sufficient conditions to
guarantee convergence of the CL method to the correct results.

An extension of these results from polynomial actions
to the more general case of an analytic $ S $  requires a more
detailed understanding of the solutions $f(x,y,\tau)$ of
$T$. Also of interest would be similar results for  compact
manifolds. Work is currently progressing in these directions.

\vskip 1cm
{\bf Acknowledgement:}
S.L. would like to thank Wes Petersen of the IPS-ETH Zurich
for illuminating discussions.
Both authors are grateful to John Klauder of the University of Florida
 for many helpful suggestions and reading the manuscript.
S.L. also wishes to thank the University of Graz and IPS-ETH Zurich for
their kind hospitality during the time in which this work was completed.
H.G. thanks Bernd Thaller for many useful discussions and his help on
the complex frequency harmonic oscillator.

\newpage


\newpage
\begin{thebibliography}{1234}
\newcommand{\authors}[1]{#1, }
\newcommand{\authorsp}[1]{{\bf #1},}
\newcommand{\journal}[1]{#1}
\newcommand{\volume}[1]{{\bf #1}}
\newcommand{\myyear}[1]{(#1)}
\newcommand{\page}[1]{#1}
\newcommand{\mytitle}[1]{}
\newcommand{\keywords}[1]{}
\newcommand{\kw}[1]{}
\newcommand{\bibi}[1]{\bibitem{#1}}

\bibitem{Met}
\authors{N. Metropolis et al.}
\journal{J. Chem. Phys.}
\volume{21} 		\myyear{1953} \page{1087}
\bibitem{Klaud1}
\authors{J.R. Klauder} in: {\em Recent Developments in High Energy Physics},
eds. H. Mitter and C.B. Lang (Springer, Wien, New York, 1983)
\bibitem{Klaud2}
\authors{J.R. Klauder, W.P. Petersen}
\journal{J. Stat. Phys.}
\volume{39} 		\myyear{1985} \page{53}
\bibitem{LinHir}
\authors{H.Q. Lin, J.E. Hirsch}
\journal{Phys. Rev. B}
\volume{34} 		\myyear{1986} \page{1964}
\bibitem{Schul}
\authors{H. Okamoto, K. Okano, L. Sch\"ulke, S. Tanaka}
\journal{Nucl. Phys. B}
\volume{324} 		\myyear{1989} \page{684};
\authors{K. Okano, L. Sch\"ulke, B. Zheng}
\journal{Phys. Lett. B}
\volume{258} 		\myyear{1991} \page{421};
\authors{K. Okano, L.Sch\"ulke, B. Zheng}
\journal{Siegen Preprint Si-91-8}
\bibitem{Arn}
\authors{L. Arnold}
{\em Stochastic Differential Equations} ( Wiley, New York 1974)
\bibitem{Yos}
\authors{K. Yosida}
{\em Functional Analysis}
(Springer, Berlin 1968)
\bibitem{Hoer}
\authors{L. H\"ormander}
{\em The Analysis of Linear Partial Differential Operators I, II}
( Springer, Berlin Heidelberg 1983)
\bibitem{Hay}
\authors{R.W. Haymaker, Y. Peng}
\journal{Phys. Rev. D}
\volume{41} 		\myyear{1990} \page{1269};
\bibitem{CyFroKiSi}
\authors{H.L. Cycon, R.G. Froese, W. Kirsch and B. Simon}
{\em Schr\"odinger Operators}
( Springer, Berlin Heidelberg 1987)

\end{thebibliography}
\end{document}